\documentclass[twocolumn,floatfix,superscriptaddress,longbibliography,PRL]{revtex4}
\usepackage{amsfonts,amssymb,amsmath,hyperref}
\usepackage[applemac]{inputenc}
\usepackage{multirow}
\usepackage{color}
\usepackage{dcolumn}
\usepackage{graphicx}
\usepackage{natbib}
\usepackage{bm}
\usepackage{ulem}
\usepackage{cleveref}
\usepackage{array}
\usepackage{tabularx}
\newcolumntype{Y}{>{\centering\arraybackslash}X}

\usepackage[paperwidth=210mm,paperheight=297mm,centering,hmargin=2cm,vmargin=2cm]{geometry}

\newif\ifhyper
\hypertrue
\ifhyper
\hypersetup{
citecolor = {green},
colorlinks = {true}, 
urlcolor = {blue} 
}
\fi

\newlength{\ldag}
\def\be{\begin{equation}}
\def\ee{\end{equation}}
\def\bea{\begin{eqnarray}}
\def\eea{\end{eqnarray}}
\def\bse{\begin{subequations}}
\def\ese{\end{subequations}}
\def\bc{\begin{center}}
\def\ec{\end{center}}

\allowdisplaybreaks 

\begin{document}

\title{Crumpled-to-flat transition of quenched disordered membranes at two-loop order}

\author{L. Delzescaux} 
\email{louise.delzescaux@sorbonne-universite.fr}
\affiliation{Sorbonne Universit\'e, CNRS, Laboratoire de Physique Th\'eorique de la Mati\`ere Condens\'ee, LPTMC, 75005 Paris, France}

\author{D. Mouhanna} 
\email{mouhanna@lptmc.jussieu.fr}
\affiliation{Sorbonne Universit\'e, CNRS, Laboratoire de Physique Th\'eorique de la Mati\`ere Condens\'ee, LPTMC, 75005 Paris, France}

\author{M. Tissier} 
\email{tissier@lptmc.jussieu.fr}
\affiliation{Sorbonne Universit\'e, CNRS, Laboratoire de Physique Th\'eorique de la Mati\`ere Condens\'ee, LPTMC, 75005 Paris, France}


\begin{abstract}

We  investigate the effects  of quenched  elastic disorder on the nature of the crumpling-to-flat transition of $D$-dimensional polymerized membranes using a two-loop computation near the upper critical dimension $D_c=4$. While the pure system undergoes  fluctuation-induced  first order transitions  below $D_c$ and for an embedding dimension $d<d_{c,pure}\simeq 218.2$, one observes, in presence of disorder,  the emergence of various  regions of  second order governed by a disordered stable fixed point for  $d<d_{c1}\sim d_{c,pure}$. This opens the possibility of a new universality class associated with the crumpling-to-flat transition of disordered membranes in $d=3$. 
\\

\end{abstract}

\maketitle 

\section{Introduction}

 Quenched disorder is an unavoidable feature of genuine materials that relies on the presence of impurities, structural defects, vacancies and similar factors. It is now well known to be at the origin of a wealth of phenomena including  deep changes in the properties, or even  in the very nature,  of the phase transitions that occur in these materials. As soon as one considers disorder, two situations can be roughly distinguished. The first one corresponds to a disorder coupled with the local energy density.  In this case when the clean  system undergoes a continuous phase  transition, the  Harris criterion \cite{harris74}  indicates that for an exponent $\alpha>0$ or, equivalently, an exponent $\nu < 2/D$  in a $D$-dimensional system, the disorder is relevant. In such case, the critical exponents should change and be controlled by a new, disordered fixed point satisfying the condition $\alpha< 0$ or $\nu > 2/D$. In the opposite case, the disorder is irrelevant and the critical exponents  are unchanged.  A typical example is the Ising model with randomly dilute impurities in $D=3$, see {\it e.g.} \cite{folk03}, for which a new universality class, {\it i.e.} distinct from the clean Ising one, with  dilution independent exponents,  has been well identified \cite{calabrese03e}. The second, and more severe situation, corresponds to a disorder that couples to the local order parameter.  The archetypal systems  are  the Random Field Ising and  Random $O(N)$ models, see \cite{tarjus20,rychkov20} for reviews.  In  these cases, the effects can be even more drastic as, according to the Imry-Ma argument \cite{imry75},  long-range ferromagnetism is destroyed by a random field in any dimension $D\le 2$ in the case of a discrete symmetry and in any dimension $D\le 4$ in the case of a  continuous  symmetry. Above these dimensions the physics  is controlled by a zero temperature fixed point so that  criticality is  dominated by quenched disorder -- instead of thermal -- fluctuations  potentially leading to a glassy behaviour. An analog situation is encountered in the case of  elastic manifolds  in a random environment, see {\it e.g} \cite{wiese07}.  Ultimately, the physics of this second class of quenched disordered  systems   relies on the existence of many metastable states, implying static avalanches and glassy dynamics   that invalidate conventional approaches. Only specific tools such as perturbative or nonperturbative {\sl functional} renormalization group (RG) approaches \cite{tarjus20} and supersymmetric conformal field theory \cite{kaviraj20} seem able  to catch the physics associated with  such a situation. 

In the case where the pure system undergoes first order transitions one also expects strong effects induced by disorder. Considering the critical exponent $\alpha$ to be formally equal to $1$ in all dimensions leads to consider the disorder as relevant \cite{berker93,cardy99}. It  was  first formulated  by Imry and Wortis \cite{imry79}, using a domain-wall argument close to that of Imry-Ma \cite{imry75} that, in $D=2$, any amount of  randomness coupled to the local energy should lead to a vanishing of the latent heat. This question has been further investigated through phenomenological RG arguments   by  Hui and Berker \cite{hui89}  who have shown that  the $q$-states random Potts models have no latent heat in $D\le 2$. This result has been  proven rigorously for a larger class of discrete models  by  Aizenman and Wehr \cite{aizenman89} and, more recently, in  the quantum case in \cite{greenblatt09,greenblatt10}.   Cardy and Jacobsen \cite{cardy97} have provided an explanation  for this situation through a mapping,  valid for strong first order transitions,  between the latent heat when randomness couples to the local energy  and the magnetization when randomness couples to the order parameter.  Several works   have then confirmed explicitly this statement in  the $q$-states random Potts  \cite{chen92,cardy97,chatelain98}  and in  the Blume-Capel \cite{malakis09,malakis10,Vatansever20,fytas18} models.  In $D>2$, the situation is  more complex  as a finite threshold of disorder can  be  required  to  change the nature of the transition \cite{hui89}. We  focus in this work  on a specific  situation, that of fluctuation-induced first order transitions which, contrary to generic first order ones, are expected to be controlled perturbatively,  as the  critical  fixed point is often reachable in this frame \cite{cardy96,cardy99}. There,   all situations  are encountered: persistence of the first order behaviours as in  randomly dilute spin systems with cubic anisotropies \cite{cardy96,cardy99,calabrese03d} or the emergence of a second order behaviour  for strong enough disorder, see, e.g, Abelian Higgs model \cite{boyanovsky82}.  In the case of frustrated magnets \cite{serreau03}, the introduction of quenched disorder  turns  the first order  into a second order behaviour in the non perturbative regime $\epsilon\sim 1$.  Finally,  the possibility of a disorder-induced first order transition has been proposed \cite{Li94,nowak22}. This plurality of behaviours calls for a  deeper understanding of the underlying mechanisms.

In this context, it is of particular interest  to study  the effects of quenched disorder on the nature of the crumpling-to-flat transition that occurs  in  polymerized membranes. Up to now,  the effects of disorder have been mainly probed   in the flat phase of membranes, as it is directly relevant  to  the physics of  graphene and graphene-like materials \cite{novoselov04,novoselov05,katsnelson12}. Notably, a lot of works  have  been dedicated to the search for a low-temperature  phase dominated by  disorder only \cite{nelson91,radzihovsky91b,radzihovsky92,morse92a,morse92b,gornyi15,coquand18,ledoussal18,saykin20b}. Recently, such a  phase has been identified, first within  a nonperturbative context \cite{coquand18}   then perturbatively at two- and three- loop order \cite{coquand20b,metayer22d}. These  approaches  have  confirmed the possibility of a whole glassy phase at low temperatures in quenched disordered membranes. More than that,  it  has been shown in \cite{coquand20}, on the basis of the investigations  performed in \cite{coquand18}, that the various scaling laws observed experimentally by Chaieb {\it et al.} \cite{chaieb06, chaieb08,chaieb13,chaieb13bis} on  partially polymerized lipid membranes were  qualitatively and quantitatively explained. Besides, the effects of disorder on the crumpling-to-flat transition of polymerized membranes have been less studied. One of the main reasons is that, in  real-life membranes,  self-avoidance is expected to destroy the crumpled phase, see \cite{gompper04}, which makes this study somewhat  formal.  However  it has been suggested, notably by  Radzihovsky and Le Doussal \cite{radzihovsky92},  then  by  Mori and Wadati \cite{mori94b}  using  Gaussian variational approximations, that self-avoiding polymerized  membranes with short-range disorder  could lie in a crumpled phase, in embedding dimension $d>2$, in contrast  with pure self-avoiding polymerized  membranes.  This has been alternately confirmed  \cite{mori96a,mori96b}  and contradicted \cite{grest90} by numerical simulations.

In this  article, we study  the effects of elastic quenched disorder on non-self avoiding  -- phantom -- $D$-dimensional polymerized membranes perturbatively at two-loop order in the vicinity of the upper critical dimension $D_c=4$. Such a computation is possible thanks to the trick introduced in \cite{delzescaux23} that allows us to deal with  derivative field theories with almost the same level of difficulty as derivative-free field theories. One derives the RG equations and studies them; one then finds,  when varying  the embedding dimension $d$,  a surprisingly rich structure that displays   whole regions  controlled by a  disorder-induced fixed point that alternate  with regions of fluctuation-induced first order transitions. The regions of second order are associated with a new universality class for membranes in the vicinity of the upper critical dimension that could extend to physical, $d=3,D=2$, membranes.

\section{Crumpling transition in pure polymerized membranes}

 One considers a generic $D$-dimensional membrane embedded  in a $d$-dimensional Euclidean space. A point in the membrane is identified inside it with the help of its local coordinates  ${\bf x}\in \mathbb{R}^D$  and in the Euclidean space with the help of  the mapping ${\bf x}\to  {\bf R}({\bf x})$ where ${\bf R}$  is a field in $\mathbb{R}^d$. The energy of a polymerized membrane has two contributions: a bending energy, common with fluid membranes, and an elastic energy, which arises from its fixed connectivity. 
 This total energy can only depend on variations  of position field $\bf R$  so that the action should display a shift symmetry ${\bf R}({\bf x})\to{\bf R}({\bf x}) + C$  where C is a
constant vector. As a consequence it expresses only in terms of  the derivatives of ${\bf R}$. Then, the relevant action is given by~\cite{paczuski88}:
\begin{align}
\begin{split}
S[\bm R] = \int  \text{d}^Dx  \ \bigg\{ &\frac{\kappa}{2} ({\partial_\alpha^2}{\bf R})^2 + \frac{{r}}{2} ({\partial_\alpha}{\bf R})^2 +  \\ 
&  \hspace{-0.2cm}+\frac{\lambda}{8} g_{\alpha \alpha}^2 + \frac{\mu}{4}\,g_{\alpha \beta}^2 \bigg\} 
\label{actionpure}    
\end{split}
\end{align}
where $g_{\alpha \beta}=\partial_\alpha {\bf R}.\partial_\beta{\bf R}$. In expression (\ref{actionpure}), Greek indices range from $1$ to $D$ and repeated indices imply summation. The bending rigidity constant is denoted by $\kappa$, while $r$ represents a tension coefficient that conveys the main temperature dependence. The -- Lam\'e -- coefficients $\lambda$ and $\mu$,  associated with quartic interactions, embody the elasticity and shear properties of the membrane. Stability  implies that $\kappa$, $\mu$, and the bulk modulus $B=\lambda+2 \mu/D$ must  be all  positive.  Action (\ref{actionpure})  is relevant to study the crumpling-to-flat transition \cite{paczuski88}: varying the tension coefficient $r$, one gets a transition between a disordered, crumpled, phase at high temperatures and an ordered, flat, phase at low temperatures with non vanishing average value of the tangent vector fields ${\bf t}_{\alpha} = \partial_{\alpha} {\bf R}$. 

The crumpling-to-flat  transition of pure polymerized membranes has  been investigated through several methods. These include $\epsilon$-expansion near the upper critical dimension $D_c=4$ at one-loop order~\cite{paczuski88}, non perturbative RG  computations  \cite{kownacki09,braghin10,hasselmann11,essafi11,essafi14,coquand18}, large-$d$ expansion~\cite{david88,aronovitz89} and self-consistent screening approximation (SCSA)~\cite{ledoussal92,ledoussal18}. In \cite{delzescaux23}, Delzescaux {\it et al} have  studied the crumpling-to-flat  transition within  the perturbative RG up to two-loop order with the help of a method based on auxiliary fields that allows to get rid  of  explicit derivatives. They have determined the critical dimension -- $d_{c,pure}(\epsilon)=218.20 - 448.25\epsilon$ -- above which one should  observe a second order phase transition controlled by a stable fixed point and below which  the absence of such a fixed point leads to expect  fluctuation-induced first order phase transitions.  For this reason, examining the impact of quenched disorder on the characteristics and nature of the crumpling-to-flat phase transition is of particular interest notably to test  if a rounding of the first order transitions is observed. 

\section{Quenched elastic disorder in polymerized membranes}
One now introduces elastic disorder in the model. Physically,  this kind of disorder corresponds  to random local inclusions and defects. Technically  they are taken into account  through  a random tensor field, $\sigma_{\alpha\beta}({\bf x})$,  that  couples to the strain tensor $g_{\alpha \beta}$:
\begin{align}
\begin{split}
S[\bm R] = \int & \text{d}^Dx \ \bigg\{ \frac{\kappa}{2} ({\partial_\alpha^2}{\bf R})^2 + \frac{{r}}{2} ({\partial_\alpha}{\bf R})^2 +  \\ 
&  \hspace{-0.5cm}+\frac{\lambda}{8} g_{\alpha \alpha}^2 + \frac{\mu}{4}\,g_{\alpha \beta}^2 -{1\over 2}\sigma_{\alpha\beta}({\bf x}) g_{\alpha \beta}\bigg\} 
\label{actionquenched}    
\end{split}
\end{align}
where the stress tensor $\sigma_{\alpha\beta}$  is chosen to be short-ranged, Gaussian, with zero mean value:
\begin{align}
\begin{split}
[\sigma_{\alpha\beta}({\bf x})\sigma_{\gamma\eta}({\bf y})]=(\Delta_{\lambda} \delta_{\alpha\beta}\delta_{\gamma\eta}+2 \Delta_{\mu} I_{\alpha \beta \gamma \eta}) \delta^{(D)}({\bf x-y})
\label{variance}    
\end{split}
\end{align}
with $I_{\alpha \beta \gamma \eta}= \frac{1}{2}(\delta_{\alpha \gamma} \delta_{\beta \eta} + \delta_{\alpha \eta} \delta_{\beta \gamma})$ where  $\alpha, \beta, \gamma$ and $\eta$ range from $1$ to $D$. In (\ref{variance})  $[...]$ denotes disorder average  while  $\Delta_{\lambda}$ and $\Delta_{\mu}$ are the variances  associated with the elastic  disorder.  The decomposition of the stress tensor $\sigma_{\alpha\beta}$ into   scalar and a symmetric traceless part with positive variance, see \cite{morse92b}, implies, in addition to the previous stability  conditions, that  $\Delta_{\mu}\ge 0$ and $B_{\Delta}=\Delta_{\lambda}+2 \Delta_{\mu}/D\ge 0$. Disorder average is performed through the replica trick  \cite{edwards75}: one introduces $n$ replica and perform the integration over the field $\sigma_{\alpha\beta}({\bf x})$ which leads to
\begin{align}
\begin{split}
S[\{ {\bf R}^A\}] &= \int  \text{d}^Dx \ \bigg\{ \frac{1}{2}\kappa ({\partial_\alpha^2}{\bf R}^A)^2 + \frac{{r}}{2} ({\partial_\alpha}{\bf R}^A)^2\\ 
&  \hspace{0.5cm}+\frac{1}{8}(\lambda \delta_{AB}-\Delta \lambda J_{AB}) g_{\alpha \alpha}^A~g_{\beta \beta}^B\\
&\hspace{0.5cm} + \frac{1}{4}(\mu \delta_{AB}-\Delta \mu J_{AB})g_{\alpha \beta}^A~g_{\alpha \beta}^B \bigg\} 
\label{actionreplica}    
\end{split}
\end{align}
with  $g_{\alpha \beta}^A=\partial_\alpha {\bf R}^A.\partial_\beta{\bf R}^A$  where capital letters are associated with the $n$ replica, $J_{AB}$ is a matrix in which all elements are equal to $1$. Summation over replica indices is implied and the limit $n\to 0$ should be taken at the end. \\

To investigate action (\ref{actionreplica})  one  has recourse to  the auxiliary fields method introduced  in \cite{delzescaux23} to study the pure case, see also \cite{ledoussal23}. One  reparametrizes the action (\ref{actionreplica}) in terms of  $2Dn$ auxiliary $d$-components fields $\{\bm A_{\alpha}^E\}$ and {$\{\bm B_{\beta}^{E'}\}$ with $\alpha,\beta=1\dots D$ and $C,E,E'=1\dots n$ so that the partition function reads:
\begin{align}
\begin{split}
    Z=\displaystyle\int &\displaystyle\prod_{\{C,E,E'\}=1}^n {\cal D}\bm R^C \hspace{-0.1cm} \displaystyle\prod_{\{\alpha,\beta\}=1}^D {\cal D}\bm A_{\alpha}^E\,  {\cal D}\bm B_{\beta}^{E'} \  e^{\displaystyle -S[\{\bm A_{\alpha}^E\}]}  \\
    &\ \times e^{\displaystyle  -i  \int d^Dx\,  \bm B_{\alpha}^E.(\bm A_{\alpha}^E-\partial_{\alpha} \bm R^E)} \, .
\end{split}
\label{actionreplicabis} 
\end{align}
Within this approach  the theory is free of field-derivatives apart from a standard kinetic term and is thus   significantly simpler than the one with the original parametrization (\ref{actionreplica}). Moreover, the fields $\bm B_{\alpha}^E$ and $\bm R^E$ appear within an expression which is quadratic in the fields in (\ref{actionreplicabis}}). As a consequence there is no interaction vertex with  legs associated with  these fields and  they do not renormalize.  Only the auxiliary fields $\{\bm A_{\alpha}^E\}$   renormalize nontrivially.

 \section{Crumpled-to flat transition of disordered polymerized membranes}
 
One has derived the two-loop (three-loop for the anomalous dimension) RG equations of model (\ref{actionreplicabis}) within the modified minimal subtraction scheme. The diagrammatics are  the same as those obtained in \cite{delzescaux23}, except that there is an additional tensorial structure  associated with replica indices.  These equations are expressed in terms of dimensionless renormalized quantities, which are defined as $g=k^{\epsilon} Z^{-2}  Z_{g} \kappa^2 g_R$ for $g \in$ $\{\lambda, \mu,\Delta_{\lambda}$,$\Delta_{\mu}\}$ and  $r=k^{2} Z^{-1}  Z_{r} \kappa r_R$ for $r$. Here, $Z$ represents the field renormalization and is defined as $\bm A_{\alpha}= Z^{1/2} \kappa^{-1/2} \bm A_{\alpha R}$, where $\bm A_{\alpha R}$ is the renormalized field,  $Z_{g}$  and $Z_r$ are  the coupling constant and tension renormalizations,  $k$  is the running  momentum scale and $\epsilon=4-D$. The RG flow of the renormalized coupling constants and tension at fixed bare theory is defined as $\beta_{g_R}=\partial_t g_R$ and $\beta_{r_R}=\partial_t r_R$ while  the running field anomalous dimension is given by $\eta=-\partial t \log Z$  with $t=\log \bar k$ where $\bar k^2 =4\pi e^{-\gamma_E} k^2$  and  $\gamma_E$  the Euler constant.   For simplicity, we  omit the index $R$ for renormalized quantities.  The expression of the RG equations, as well as that of the exponent $\nu$ are too long  to be displayed within the main text and are provided in the appendix. We indicate  that the first nonvanishing contribution to the anomalous dimension occurs at three-loop order. We have verified that  the RG equations  $\beta_X(\lambda,\mu,\Delta\lambda,\Delta\mu,d)$ with $X=\lambda,\mu,\Delta\lambda,\Delta\mu$ and  $\beta_r(\lambda,\mu,\Delta\lambda,\Delta\mu,r,d)$ at two-loop order as well as  $\eta(\lambda,\mu,\Delta\lambda,\Delta\mu,d)$ at three-loop order reproduce, when disorder vanishes,  the equations derived in \cite{delzescaux23}. On the other hand, one has checked  nontrivial identities at $d=0$  that generalize those encountered  in the context of spin systems with quenched disorder for a vanishing number of the order parameter components (which is the embedding  dimension $d$ here),  see  
\cite{calabrese04b}. One has:
\begin{equation}
\begin{array}{ll}
\beta_X(\lambda,\mu,\Delta\lambda,\Delta\mu,0)&-\ \beta_{\Delta X}(\lambda,\mu,\Delta\lambda,\Delta\mu,0)\\
\\
&=\beta_X(\lambda-\Delta\lambda,\mu-\Delta\mu,0,0,0)
\end{array}
\nonumber
\end{equation}
with $X=\lambda,\mu$, then
{\begin{equation}
\begin{array}{ll}
&\beta_r(\lambda,\mu,\Delta\lambda,\Delta\mu,r,0)=\beta_r(\lambda-\Delta\lambda,\mu-\Delta\mu,0,0,r,0)
\end{array}
\nonumber
\end{equation}
and for the  anomalous dimension :
\begin{equation}
\eta(\lambda,\mu,\Delta_{\lambda},\Delta_{\mu},0)=\eta(\lambda - \Delta_{\lambda},\mu - \Delta_{\mu},0,0,0)\ . 
\nonumber
\end{equation}

\subsection{Disorder-free case}

Before considering the model studied here, let us first recall the main features of disorder-free model \cite{paczuski88}. In this case there exist four fixed points among  which two -- the Gaussian one and  another one -- are always unstable and play no role in $D<4$. The properties and stability of the two remaining fixed points depend on the embedding dimension $d$. For $d>d_{c,pure}$  one fixed point, which  we call  FP$_1$,  is stable and controls the crumpling-to-flat transition while  the other one, FP$_2$, is unstable. When $d$ decreases and  reaches  $d_{c,pure}$,   FP$_1$ and FP$_2$ get complex with conjugate coordinates. In the absence of any physical stable fixed point the transitions are  expected to be fluctuation-induced first order ones. The curve $d_{c,pure}(\epsilon=4-D)$ separating the second order region for $d>d_{c,pure}$ and the first order one for $d<d_{c,pure}$ has been obtained at next to leading order in  $\epsilon$ within a two-loop order computation  \cite{delzescaux23} and is given by: $d_{c,pure}(\epsilon)=218.20-448.25\, \epsilon$, see  Fig.\ref{dc1dc2pure}.

\subsection{Disordered case}

Let us now consider the effects of disorder. A computation of $\nu-2/D$ from the expression obtained for $\nu$ in the pure case \cite{delzescaux23} shows that  this quantity  is positive for  $d>d_{c1}$ and  negative for  $d<d_{c1}$ with $d_{c1}=218.29$ in the vicinity of $D=4$.  Thus one expects a  change of critical behaviour due to the disorder in this last  case.
With this in mind one  investigates the RG equations in the  presence of disorder at  one and two-loop orders. As the flow is too complicated, even at one-loop order, to be studied analytically in $\epsilon$, one 
solves it numerically. 

\subsubsection{One-loop order}

At one loop order, one identifies three nontrivial fixed points respecting the energy  stability criteria and playing an important role: FP$_1$ and  FP$_2$,  that were already present in the disorder-free case and a, new, disordered fixed point, FP$_D$. Note that FP$_2$  is always unstable.  As in the disorder-free case, the  dynamical stability of the fixed points  depends on the value of the embedding dimension $d$. Specifically, as far as  $d$ is above  a critical value  $d_{c1}=218.29$, see  Fig.\ref{dc1dc2pure},  FP$_1$ is found to be stable and  FP$_D$  unstable in agreement with the considerations above. When $d$ falls below $d_{c1}$, the fixed point FP$_1$ loses its stability for the benefit of FP$_D$ that thus controls the crumpling-to-flat transition. As for the  fixed points FP$_1$ and FP$_2$, they get complex coordinates for the same critical value as in the pure case, $d_{c,pure}=218.20$, leaving  FP$_D$ alone and stable, see Fig.\ref{dc1dc2pure}. This situation persists when decreasing $d$ until it reaches the  value $d_{c,dis1}=74.00$ at which  FP$_D$ becomes, again,  unstable, see Fig.\ref{dc1dc2},  so that  the transition is again expected to be fluctuation-induced first order. (Note that  for  $d_{c2}=218.25<d<d_{c1}$,  FP$_D$ is a stable node, while   for $d<d_{c2}$ the eigenvalues at  FP$_D$ are complex  implying that  it is a focus fixed point. This fixed point is surrounded by an unstable limit cycle,  a behaviour  that  has been already observed in various  contexts involving quenched disorder: $\phi^4$ theory  with long-range-correlated quenched disorder \cite{weinrib83},  impure Abelian  Higgs model or superconductors \cite{boyanovsky82,athorne85,athorne86}). 

\begin{figure}[h!]
\includegraphics[scale = 0.55]{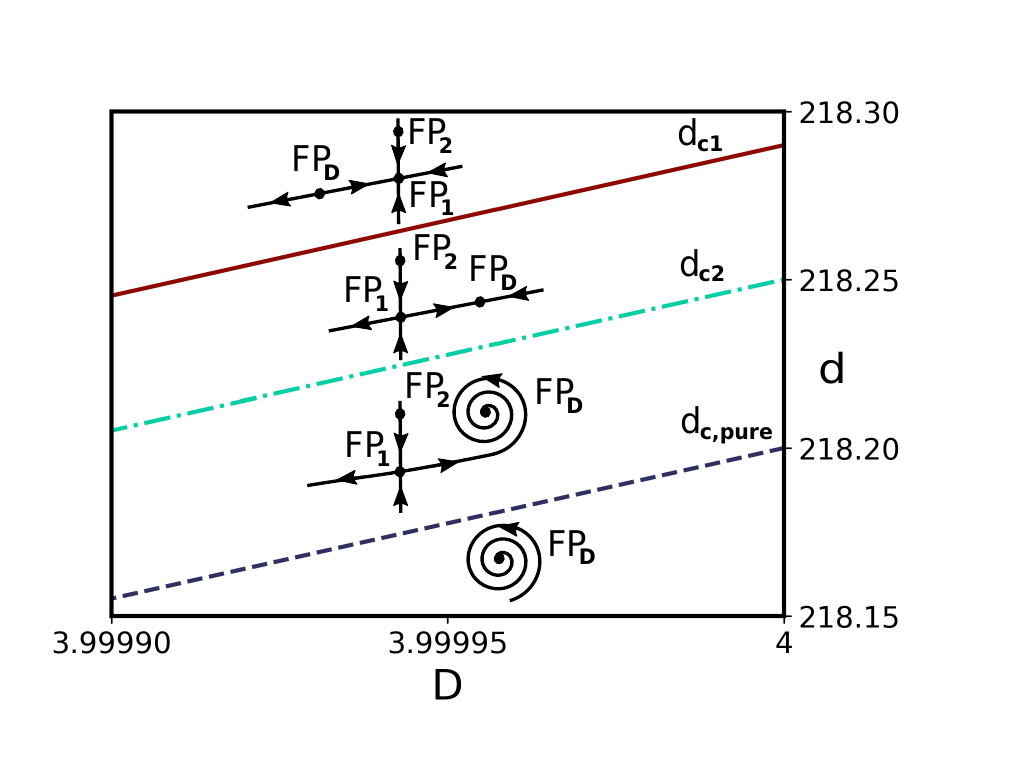}
\caption{The solid line $d_{c1}$  separates the region $d>d_{c1}$ controlled by the  disorder-free stable  fixed point FP$_1$  and the region $d<d_{c1}$ controlled by  the  disordered stable fixed point FP$_D$. The dash-dotted line $d_{c2}$  separates the region $d>d_{c2}$  where the fixed point FP$_D$ is a stable node and the region $d<d_{c2}$ where it becomes a stable focus. Finally the dashed line $d_{c,pure}$ delimits the region $d>d_{c,pure}$ in which   FP$_1$ and FP$_2$ are real. This last curve is the same as in the pure case. The RG flows are schematically described in the different regimes of $d$.}
\label{dc1dc2pure}
\end{figure}

A rather surprising finding is that, when $d$ is further decreased below $d_{c,dis1}$,  a new second order region emerges at $d_{c,dis2}=14.69$,  controlled by the stable focus fixed point FP$_D$, see Fig.\ref{dc1dc2}. This region extends down to  $d_{c,dis3}^{(1)}=5.56$ where  FP$_D$  gets complex coordinates. A detailed  analysis shows that, at the same time, the behaviour of the corresponding coupling constants at this order is linear in $\epsilon$ and singular as a function of $d-d_{c,dis3}^{(1)}$. One has typically $g^*\sim \epsilon/(d-d_{c,dis3}^{(1)})$ for   $g \in$ $\{\lambda, \mu,\Delta_{\lambda}$,$\Delta_{\mu}\}$; see below for a discussion.   

Finally,  there exists, in the pure case,  a fixed point at  $d=1$ \cite{safari22,ledoussal23}; it is, in fact,  found everywhere in the  embedding dimensions range $[0,d_1=1.12]$. Also one finds that this fixed point is destabilized by the disorder in this whole range of dimensions.  According to the values of the initial disorder, the flow  shows either  a runaway  at infinity, the sign of first order transitions, or drives  the system  towards a completely  disordered fixed point FP$_{CD}$   with $\lambda=\mu=0$ and $\Delta \mu=-\Delta \lambda$ with $\Delta\mu<0$ and $\Delta\lambda+2 \Delta\mu/D>0$ in the vicinity of $D=4$. The identification of its   precise nature needs further investigations and will be  discussed in \cite{mouhanna24}.

The conclusion of  this  leading order RG analysis is that  elastic disorder in polymerized membranes smoothens  the transition from first order to a second order one in the ranges of values  $d\in  [74.00,218.29]$  and $d\in [5.56,14.69]$ in  the neighborhood  of $D=4$ dimensions. For $d\in [0,1.12]$  it  also destabilizes the stable fixed point associated with  the pure system but in a more elusive way.

\begin{figure}[h!]
\includegraphics[scale = 0.6]{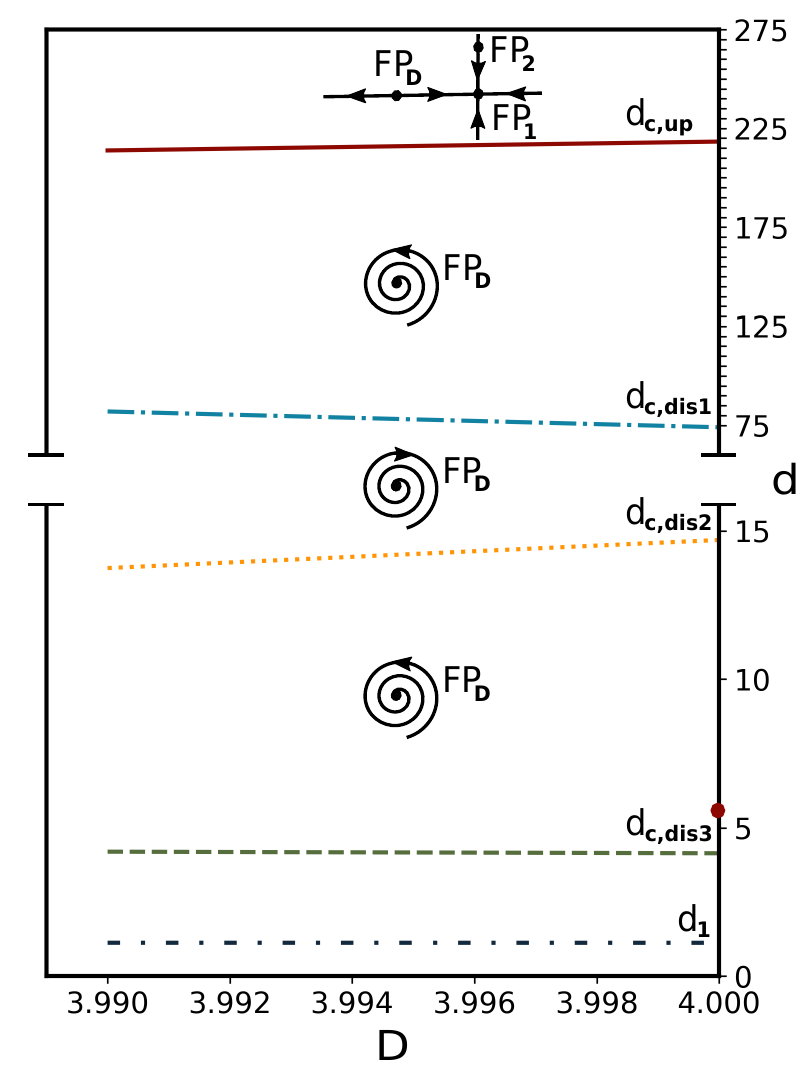}
\caption{The set of  lines $d_{c,up}\equiv \{d_{c1},d_{c2},d_{c,pure}\}$   (which coincide at the scale of the figure) separates the region $d>d_{c,up}$ controlled by the  disorder-free stable  fixed point FP$_1$ and the region $d<d_{c,up}$ controlled by  the  disordered stable fixed point FP$_D$. The  dash-dotted   line  $d_{c,dis1}$ separates the region $d>d_{c,dis1}$ controlled by  a disordered stable focus fixed point FP$_D$  from the region $d_{c,dis2}<d<d_{c,dis1}$  without any stable fixed point. Below the dotted line $d_{c,dis2}$, one encounters a  region controlled again by FP$_D$ until $d$ reaches the value $d_{c,dis3}$. At first order $d_{c,dis3}$ is  represented by a point at $d=5.56$ and, at two-loop order by the  dashed line (see the text for discussion). Below this line, there is no stable fixed point until one reaches the dash-dotted line $d_1$ under which  a stable fixed point reappears in the pure case; it is however destabilized by the disorder.}
\label{dc1dc2}
\end{figure}

\subsubsection{Two-loop order}

At two-loop order, the overall qualitative picture found at one-loop order is essentially unchanged: one  recovers the same fixed points and behaviours with, alternatively, second order and first order regions, as shown in Fig.\ref{dc1dc2}. At this order, one  has now access to the linear contributions in $\epsilon$ to $d_{c1}$, $d_{c2}$  as well as  $d_{c,dis1},d_{c,dis2}$, $d_{c,dis3}$  and $d_1$, which  were anticipated in both Fig.\ref{dc1dc2pure} and  Fig.\ref{dc1dc2}.  By varying $\epsilon$ at order $10^{-7}$, one  recovers  the result of the pure case \cite{delzescaux23},  given by the line $d_{c,pure}(\epsilon)=218.20 - 448.25\, \epsilon$. Moreover, one gets the equation of the lines  $d_{c1}(\epsilon)=218.29 - 448.10\, \epsilon$,   $d_{c2}(\epsilon)=218.25 - 448.18\, \epsilon$, see Fig.\ref{dc1dc2pure}, $d_{c,dis1}(\epsilon)=74 + 799.46\,\epsilon$, $d_{c,dis2}(\epsilon)= 14.69 - 94.72\, \epsilon$, see  Fig.\ref{dc1dc2}. Finally, one gives the  curve $d_1(\epsilon)=1.12 + 0.14 \, \epsilon$ which  gives the upper limit of the domain where  a fixed point reappears in the pure case.

A striking  fact emerges from  the study of the curve $d_{c,dis3}(\epsilon)$.  Specifically, its  coefficient of order $\epsilon^0$ is modified by the two-loop order contribution, going from $d_{c,dis3}^{(1)}(\epsilon=0)=5.56$ to $d_{c,dis3}^{(2)}(\epsilon=0)=4.14$. An important observation is that, at this order,  the behaviour of the coupling constants in the vicinity of  $d_{c,dis3}$  is now regular  as a function of $d-d_{c,dis3}^{(2)}$  and  exhibits a square-root  dependence on  $\epsilon$ as far as one is very close to   $d_{c,dis3}^{(1)}(\epsilon=0)=5.56$. Below this value one encounters a behaviour that is no longer perturbative in $\epsilon$ and has thus no physical meaning. Such features  are  extremely reminiscent of what happens in the context of randomly dilute $m$-vector model where the coupling constants behave, at one-loop order, as $\epsilon/(m-1)$  and,  at two-loop order, as  $\sqrt \epsilon$ in the strict vicinity of $m=1$ \cite{khmelnitskii75,grinstein76}. One thus identifies here two extremely analogous phenomena, with the difference that limit where the dimension reaches the critical  dimension $d_{c,dis3}^{(1)}(\epsilon=0)=5.56$, at which the stable fixed point disappears, assumes the role of  the   $m=1$   limit in the context  of  the randomly dilute $m$-vector model. The conclusion of this discussion is that  the value obtained at two-loop order, $d_{c,dis3}^{(2)}(\epsilon=0)=4.14$,  and  its extension to finite values of $\epsilon$, $d_{c,dis3}(\epsilon)=4.14 +5.32 \, \epsilon$,  are very questionable.

Finally, as for the $d<d_1(\epsilon)$  case, the two-loop order confirms that the  disorderless fixed point is destabilized by  disorder.  Moreover one observes that, at this  order,  the flow  it attracted by the fixed point  FP$_{CD}$ for {\sl all}  initial conditions.

Several remarks are in order. First, the lines $d_{c,pure}(\epsilon)$,  $d_{c1}(\epsilon)$, $d_{c2}(\epsilon)$ and  $d_{c,dis2}$  behave (decrease)  with $\epsilon$  in an extremely similar way, as shown in Fig.\ref{dc1dc2pure} and  \ref{dc1dc2}.  This is in strong contrast  with  the lines $d_{c,dis1}(\epsilon)$,   $d_{c,dis3}(\epsilon)$  and $d_1(\epsilon)$   which are  increasing functions  of $\epsilon$, an unusual behaviour. As noted  previously, the expression of  $d_{c,dis3}(\epsilon)$ is very doubtful. One focuses on $d_{c,dis1}(\epsilon)$ which displays a second unusual behaviour:  the series  $d_{c,dis1}(\epsilon)$  is characterized by a coefficient for  the order epsilon that is  10  times higher than that of the  leading order. Let  us recall that for most  series occurring in similar  contexts,  here  the lines  $d_{c,pure}(\epsilon)$, $d_{c1}(\epsilon)$ and $d_{c2}(\epsilon)$, the lines $N_c(\epsilon)=21.8-23.4\, \epsilon$ in  frustrated magnets \cite{jones76,bailin77}, or  $N_c(\epsilon)=365.9-641.057\, \epsilon$ in scalar electrodynamics, see \cite{ihrig19} and references therein, or   $N_c(\epsilon)=718-990.83\,  \epsilon$  for electroweak transition \cite{arnold94},  one observes  that  the  coefficients for the orders  $\epsilon^0$ and $\epsilon^1$ are  of the  same order of magnitude. In the context of frustrated magnets in the presence of quenched disorder,  a curve analogous to  $d_{c,dis1}(\epsilon)$, in the sense that it restricts the region of a  stable fixed point with disorder,  has been obtained in \cite{serreau03}.  However it takes the same form as $N_c(\epsilon)$ and, in fact, strictly {\sl identifies} with $N_c(\epsilon)$ at first order in $\epsilon$:  $N_{c2}(\epsilon)=21.8 - 23.4\, \epsilon$.  All these facts underline the singularity of the situation  and make nontrivial  the question of the extension of the second order region controlled by the fixed point with disorder  FP$_D$. It is clear that higher-orders perturbation theory, involving series with higher, up to five or six loop,  orders in $\epsilon$, together with resummation procedure or, even, nonperturbative approaches  are needed to clarify this situation and notably the physics in D = 2.

\section{Conclusion}
Employing the recently developed auxiliary fields method,  we have investigated the influence of quenched elastic disorder on the crumpling-to-flat transition of  polymerized membranes at two-loop order. We have observed the appearance of  a new, disordered, stable fixed point governing  second order phase transitions in several regions in the $(d,D)$ plane.  These regions should be further investigated; however our approach shows that elastic disorder has the expected effect of rounding first order phase transitions in membranes and could lead to a new universality class for physical, $(d=3,D=2)$ membranes. For $d$ in the vicinity of $d=1$, the disorder induces either first order transtions -- what is rather unusual --  or convergence towards a fully disordered fixed point. The present  study also  reveals  various other interesting behaviours:  nontrivial bifurcations within the RG flow \cite{gukov17}   such as  the occurrence of  focus fixed points accompanied by a limit cycle induced by  disorder, a phenomenon  observed in both classical and quantum cases, see e.g. \cite{yerzhakov21} and reference therein. 
Finally, for a specific critical dimension one  gets an  analytical  behaviour  extremely  analogous to  that   encountered in the context of  the randomly dilute $m$-vector model but with a very  different physical meaning.  Membranes with elastic disorder thus shows   an unsuspected richness that should be further investigated at higher order or, even more, nonperturbatively \cite{kownacki24}. 

\acknowledgements

L.D. and D.M.  greatly thank J.-P. Kownacki  for discussions.

\appendix

\begin{widetext} 

\section{Renormalization group equations at two-loop order} 
\label{app1}
\nonumber

We give here the RG equations  at two-loop order for the dimensionless coupling constants and at first nontrivial, three-loop, order for the anomalous  dimension with $c_1 = 1/96\pi^2$. 
\begin{align}
\begin{split}
& \beta_{\lambda}(\lambda,\mu,\Delta\lambda,\Delta\mu,d) =-\epsilon \,\lambda + c_1\, \bigl(-8 \, (\Delta_{\lambda} + 4 \, \Delta_{\mu})\, \lambda + (7 + 6\,d)\, \lambda^2 + \bigl((34  + 6\,d)\lambda - 4\, \Delta_{\lambda} - 16\,  \Delta_{\mu}\bigl)\, \mu + (15 + d)\, \mu^2\bigl)\\
&+ \frac{c_1^2}{6}\, \Bigl(-130\,\lambda\,\Delta_{\lambda}^2 - 748\, \lambda\, \Delta_{\lambda}\, \Delta_{\mu} + 282\, \lambda\, \Delta_{\mu}^2 + 3\, (55 + 12\, d)\, \lambda^2\, \Delta_{\lambda}  + (569 - 114\, d)\, \lambda^2\, \Delta_{\mu} - (52 + 69\,d)\, \lambda^3\\
&+103\, \mu\, \Delta_{\lambda}^2+ 514\,\mu\, \Delta_{\lambda}\, \Delta_{\mu}  + 591\, \mu\, \Delta_{\mu}^2 + 
2\, \bigl((316 - 61\, d)\, \Delta_{\lambda} + (-206 + 165\, d)\, \Delta_{\mu}\bigl)\, \lambda\, \mu + \bigl(-541 + (16 - 54\, d)\, d\bigl)\, \lambda^2\, \mu\\
&- 353\, \mu^2\, \Delta_{\lambda} - 88\, d\, \mu^2\, \Delta_{\lambda} - 577\, \mu^2\, \Delta_{\mu} + 150\, d\, \mu^2\, \Delta_{\mu} - \bigl(-110 + (281 + 36\, d)\,d\bigl)\, \lambda\, \mu^2 + \bigl(95 - 2\,(56 + 3\, d)\, d\bigl)\, \mu^3\Bigl)\\
\end{split}
\end{align}
\begin{align}
\begin{split}
&\beta_{\mu}(\lambda,\mu,\Delta\lambda,\Delta\mu,d)=-\epsilon\, \mu + c_1\, \bigl((\lambda -2\, \Delta_{\lambda} - 8\, \Delta_{\mu})\, \lambda - 10\, (- \lambda + \Delta_{\lambda} + 4 \Delta_{\mu})\, \mu + (21 + d)\, \mu^2\bigl) + \frac{c_1^2}{12}\, \bigl(103\, \lambda\, \Delta_{\lambda}^2\\
&+ 514\, \lambda\, \Delta_{\lambda}\, \Delta_{\mu} + 591\, \lambda\, \Delta_{\mu}^2 - 3\, (55 + 96\, d)\, \lambda^2\, \Delta_{\lambda} - (263 + 372\, d)\, \lambda^2\, \Delta_{\mu} + (55 + 96\, d)\, \lambda^3 -157\, \mu\, \Delta_{\lambda}^2 + 1155\, \mu\, \Delta_{\mu}^2 \\
&-2\, (199 + 120\, d)\, \lambda\, \mu\, \Delta_{\mu} + (289 + 470\, d)\, \lambda^2\, \mu - (578\, \lambda + 982\, \Delta_{\mu}  + 940\, d\, \lambda)\, \mu\, \Delta_{\lambda} -\mu^2\, \Delta_{\lambda}\, (421 + 146\, d)\\
&- 1151\, \mu^2\, \Delta_{\mu} + 858\, d\, \mu^2\, \Delta_{\mu} + 421\, \lambda\, \mu^2 + 146\, d\, \lambda\, \mu^2 + (475-212\, d)\, \mu^3  \bigl)\\
\end{split}
\end{align}
\begin{align}
\begin{split}
&\beta_{\Delta_{\lambda}}(\lambda,\mu,\Delta\lambda,\Delta\mu,d)=-\epsilon\, \Delta_{\lambda} + c_1\, \Bigl(-7\, \Delta_{\lambda}^2 + \bigl(2\, (\lambda + 7\, \mu) -15\, \Delta_{\mu}+2\, d\,(3\, \lambda + \mu)\bigl)\, \Delta_{\mu} + \bigl(6\,(\lambda+ 5\,\mu) -34\, \Delta_{\mu}\\
&+ 6\, d\, (2\, \lambda + \mu)\bigl)\, \Delta_{\lambda}\Bigl) +\frac{c_1^2}{6}\, \Bigl(-52\, \Delta_{\lambda}^3 + \bigl(-541\, \Delta_{\mu} + 2\, (13\, \lambda + 322\, \mu) + d\, (309\, \lambda + 26\, \mu)\bigl)\, \Delta_{\lambda}^2 + 110\, \Delta_{\lambda}\, \Delta_{\mu}^2\\
&- (450 + 90\, d + 108\, d^2)\, \lambda\, \mu\, \Delta_{\lambda} + (9 - 171\, d)\, \lambda^2\, \Delta_{\lambda} - 9\, (27\, + 41\, d + 4\, d^2)\, \mu^2\, \Delta_{\lambda} + 2\,\bigl(167\, \lambda + 147\, \mu\\
&- d\, (10\, \lambda - 167\, \mu)\bigl)\, \Delta_{\lambda}\, \Delta_{\mu} + (28 - 98\, d - 54\, d^2)\, \lambda^2\, \Delta_{\mu} + 95\, \Delta_{\mu}^3 - 8\, \bigl(24 + (29 + 9\, d)\, d\bigl) \lambda\, \mu\, \Delta_{\mu}\\
&- 2\, (146 + 93\, d + 9\, d^2)\, \mu^2\, \Delta_{\mu} + \bigl((172 + 67\, d)\, \lambda + 2\, (153 + 86\, d)\, \mu\bigl)\, \Delta_{\mu}^2 \Bigl)\\
\end{split}
\end{align}
\begin{align}
\begin{split}
&\beta_{\Delta_{\mu}}(\lambda,\mu,\Delta\lambda,\Delta\mu,d)=-\epsilon\, \Delta_{\mu} - c_1\, \Bigl(\Delta_{\lambda}^2 + 10\, \Delta_{\lambda}\, \Delta_{\mu} + \bigl(21\, \Delta_{\mu} - 2\,(\lambda + \mu + d\, \mu)\bigl)\, \Delta_{\mu}\Bigl) + \frac{c_1^2}{12}\, \Bigl(55\, \Delta_{\lambda}^3 + \bigl(421\, \Delta_{\mu}\\
&- 16\, (4 + 55\, d)\, \lambda - 32\, (57 + 2\, d)\, \mu\bigl)\, \Delta_{\lambda}\, \Delta_{\mu} + \bigl(475\, \Delta_{\mu}^2 + 2\, (13 + 49\, d)\, \lambda^2 + 4\, (111 + 13\, d)\, \lambda\, \mu+  2\, (137 + 111\, d)\, \mu^2\\
&+ 2\, (85 - 110\, d)\, \lambda\, \Delta_{\mu} + 2\, (-135 + 85\, d)\, \mu\, \Delta_{\mu}\bigl)\, \Delta_{\mu} + \bigl(289\, \Delta_{\mu} - (62 + 192\, d)\, \lambda - (446 + 62\, d)\, \mu \bigl)\, \Delta_{\lambda}^2\Bigl)\\
\end{split}
\end{align}
\begin{align}
\begin{split}
&\beta_{r}(\lambda,\mu,\Delta\lambda,\Delta\mu,r,d)=-2\, r +3\, c_1\, r\, \Bigl((1+2\, d)\, \lambda+
(5+d)\, \mu-\Delta \lambda-5\, \Delta\mu\Bigl)+\frac{3\, c_1^2}{4}\, \,  r\, \Bigl(\Delta\lambda^2 -  27\, \Delta\mu^2\\
&+ (1 - 19\, d)\, \lambda^2 - 2\, \bigl(25 + 5\, d  + 6\, d^2\bigl)\, \lambda\, \mu - \bigl(27 + 41\, d  + 4\, d^2\bigl)\, \mu^2 - 2\,  \bigl(25\, \Delta\mu + (1 - 13\, d)\,\lambda  + (-25 + d)\, \mu\bigl)\,\Delta\lambda\\
&+ 2\, \bigl((25 + d)\, \lambda + (27 + 25\, d)\, \mu\bigl)\,\Delta\mu \Bigl)\ . 
\end{split}
\end{align}
The equation for $r$  provides the exponent $\nu$
\begin{align}
\begin{split}
&\nu(\lambda,\mu,\Delta\lambda,\Delta\mu,d)=\frac{1}{2}+{3\, c_1\over 4}\, \Bigl((1+2\, d)\, \lambda+(5+d)\, \mu -\Delta\lambda-5\, \Delta\mu\Bigl)+{3\, c_1^2\over 16}\, \Bigl(7\,\Delta\lambda^2 + 123\,  \Delta\mu^2\\
&+ \bigl(7 + d\, (5 + 24\, d)\bigl)\, \lambda^2 + 2\, \bigl(5 + d\, (61 + 6 \,d))\, \lambda\, \mu + (123 + d\, (19 + 2\, d)\bigl)\, \mu^2 - 
 2\,  \bigl((5+ 59\, d)\,\lambda + (123 + 5\, d)\, \mu\bigl)\,\Delta\mu\\
 &+ 2\,  \bigl(5\, \Delta\mu + (-7 + d)\, \lambda - (5 + 7\, d)\, \mu\bigl)\,\Delta\lambda\Bigl)\ . 
\end{split}
\end{align} 
Finally the anomalous dimension at three loop order  is given by:
\begin{align}
\begin{split}
&\eta(\lambda,\mu,\Delta\lambda,\Delta\mu,d)={9 c_1^3}\, \biggl(-6\, \Delta_{\lambda}^3 - 76\, \Delta_{\mu}^3 + 3\,\bigl(2\, (3\, \lambda + 8\, \mu) + (5\, \lambda + 6\, \mu)\,d -16\, \Delta_{\mu}\bigl)\Delta_{\lambda}^2 + \bigl((110 + 59\, d)\, \lambda\\
&+ 2\, (114 + 55\, d)\, \mu \bigl)\, \Delta_{\mu}^2 + (2 + d)\, (\lambda + 2\, \mu) \bigl((3 + 2\, d)\, \lambda^2 + 2\, (9 + d)\, \lambda\, \mu + (19 + d)\, \mu^2\bigl) - 2\, (2 + d)\, \bigl(3\, (4 + d)\, \lambda^2\\
&+ 5\, (11 + d)\, \lambda\, \mu + 3\, (19 + d)\, \mu^2\bigl)\, \Delta_{\mu} -\Bigl ( 110\, \Delta_{\mu}^2-2\, \bigl((48 + 31\, d)\, \lambda + 2\, (55 + 24\, d)\, \mu \bigl)\, \Delta_{\mu}\\
&+ (2 + d)\, \bigl((9 + 6\, d)\, \lambda^2 + 12\, (4 + d)\, \lambda\, \mu + 5\, (11 + d)\, \mu^2\bigl) \Bigl)\, \Delta_{\lambda}\biggl).   
\end{split}
\nonumber
\end{align}

\end{widetext}


\end{document}